\documentclass[twoside,twocolumn,fleqn]{article}
\usepackage{epsf}
\usepackage{latexsym}
\usepackage{espcrc2}

\newcommand{\AmS}{{\protect\the\textfont2
  A\kern-.1667em\lower.5ex\hbox{M}\kern-.125emS}}
\newlength{\bredde}
\renewcommand{\to}{\rightarrow}
\newcommand{\beq}[1]{\begin{equation}\label{#1}}
\newcommand{\eeq}{\end{equation}}
\newcommand{\bea}[1]{\begin{eqnarray}\label{#1}}
\newcommand{\eea}{\end{eqnarray}}

\newcommand{\vev}[1]{ {\langle #1 \rangle} }
\def\slash#1{\settowidth{\bredde}{$#1$}\ifmmode\,\raisebox{.15ex}{/}
\hspace*{-\bredde}#1\else$\,\raisebox{.15ex}{/}\hspace*{-\bredde} #1$\fi}

\def\e{{\rm e}}
\def\ln{{\rm ln}}

\begin{document}
{\onecolumn

\title{Complex zeros of the $2d$ Ising model 
on dynamical random lattices \thanks{NBI-HE-97-37, NORDITA-97/47 P}}

\author{J.~Ambj{\o}rn, K.~N.~Anagnostopoulos \address{The Niels Bohr Institute,
Blegdamsvej 17, DK-2100 Copenhagen \O, Denmark}
and U.~Magnea \address{NORDITA, 
Blegdamsvej 17, DK-2100 Copenhagen \O, Denmark}\\
\vspace{1mm}
(Presented by U.~Magnea)}

\begin{abstract}
We study the zeros in the complex plane of the partition function for
the Ising model coupled to $2d$ quantum gravity for complex magnetic
field and for complex temperature. We compute the zeros by using the exact
solution coming from a two matrix model and by Monte Carlo simulations
of Ising spins on dynamical triangulations. We present evidence 
that
the zeros form simple one-dimensional patterns in the complex plane, and
that the critical behaviour of the system is governed by the scaling
of the distribution of singularities near the critical point. 
\end{abstract}

\maketitle}
\nopagebreak

\section{THE MODEL}

We study the Ising model on dynamical square and triangular lattices with the
spins located on the $N$ faces and on the $N_v$ vertices, respectively. 
The partition function for a fixed lattice $G_N$ is given by
\begin{equation}
\label{*1}
Z_{{\rm flat}}(G_N,\beta,H) =
\sum_{\{\sigma\}} \e^{{\beta}\sum_{\vev{i,j}}\sigma_i
\sigma_j + H \sum_i \sigma_i} \label{eq:ZGN}
\end{equation}
and the partition function $Z(N,\beta,H)$ for the model coupled to
quantum gravity is obtained by summing $Z_{{\rm flat}}(G_N,\beta,H)$ over
all lattices $G_N$ with spherical topology.
The system undergoes a third order phase transition \cite{BK}. 
Assuming scaling, the
critical exponents (that are simply related to the flat space exponents
\cite{KPZ}) can be computed directly from the 
equivalence of 
$Z(N,\beta,H)$ to the free energy of a hermitean two--matrix model 
\cite{BK}. In the usual notation,
$\beta = 1/2$, $\gamma = 2$, $\delta = 5$, $\nu d_H = 3$
($d_H\approx 4$ is the Hausdorff dimension of the system).

If we fix $\beta$ ($H$), $Z_{{\rm flat}}(G_N,\beta,H)$ is a
polynomial in $y\equiv \e^{-2H}$ ($c\equiv \e^{-2\beta}$).   
In the thermodynamic limit its zeros (called Lee--Yang \cite{YL} or Fisher
\cite{F} zeros respectively)
form dense sets on lines, 
which are Stokes lines for $Z$. The Lee--Yang zeros lie on the unit circle
\cite{YL} for fixed $G_N$. 
One then defines a density of zeros
$\rho_{YL}(\beta,\theta)$ ($H=i\theta$) ($\rho_F(\beta, H)$ for Fisher zeros). 
In the high temperature phase the Lee--Yang zeros form a
gap such that $\rho_{YL}(\beta,\theta) = 0$ for
$|\theta|<\theta_0$. As $\beta \to \beta_c$, the gap closes
and the zeros pinch the real axis at $y=1$,
signalling the onset of the phase transition.
 
\section{EXACT CALCULATION}

The exact solution of the Ising model on a dynamical lattice
was given in \cite{BK}. 
The grand canonical partition function $Z(g,\beta,H)$ 
is given by 
$Z(g,\beta,H)\equiv
 \sum_{N=1}^\infty k^N Z(N,\beta,H)$,
where $k\equiv -4gc/(1-c^2)^2$. 
Using the exact solution we determined 
$Z(N,\beta,H)$ for $N\le 14$. Then its   
roots for either fixed
$\beta$ or fixed $H$ were determined.

Our results indicate that the Lee--Yang zeros lie 
on the unit circle (cf. \cite{Staudacher}), 
{\it also after summing over all $G_N$}. Fig.~1 indicates a 
vanishing gap in $\rho_{YL}(\beta_c,\theta)$ as $N \to \infty$. 
We also observed the vanishing of the gap 
for a fixed lattice size as $\beta \to \beta_c$ 
from the hot phase. 

\centerline{\epsfxsize=7.0cm \epsfysize=4.67cm \epsfbox{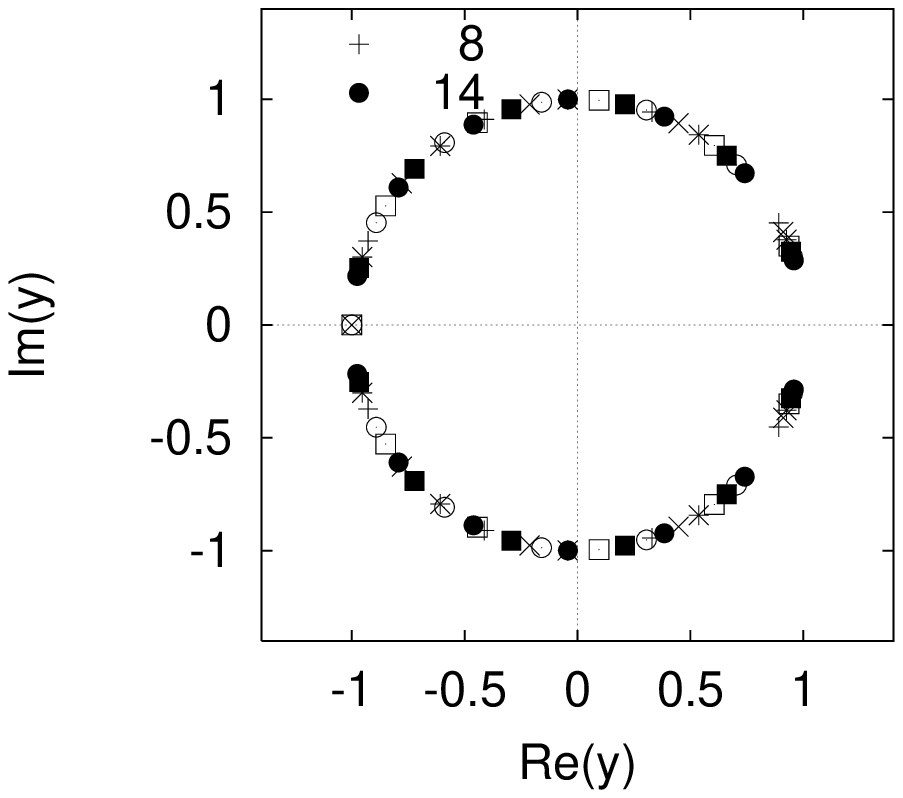}}

\vspace{-2mm}
\noindent {\bf Figure 1.} Lee--Yang zeros 
for dynamical lattices of varying size $N=8,\ 9,\ ...,\ 14$ at $\beta_c$.

\vspace{1mm}
Also the Fisher zeros form $1d$ patterns. For $H=0$ and $N\to \infty$ they 
approach the (physical and non-physical) critical points $c=\pm 1/4$. There 
are also Fisher zeros on the imaginary axis, that flow to $c=\pm i \infty$.
Fig.~2 shows the Fisher zeros for $H=0$ and $H\neq 0$ at fixed $N$.    
Mapping the zeros onto the $\tilde c$-plane, where the tilde refers to the 
{\it dual} spin model (recall the usual duality 
relation $\tilde{c} \equiv \e^{-2\tilde{\beta}} = {\rm tanh}(\beta)$), we 
observed the existence of an {\it antiferromagnetic} phase transition
in the dual model in agreement with ref.~\cite{BJ}. This AF transition
is absent in the original model where the spins are located on the faces of
the lattice. 

\centerline{\hspace{1.6cm}
\begin{minipage}[t]{6.5cm}
\centerline{\epsfxsize=6.5cm \epsfysize=4.33cm \epsfbox{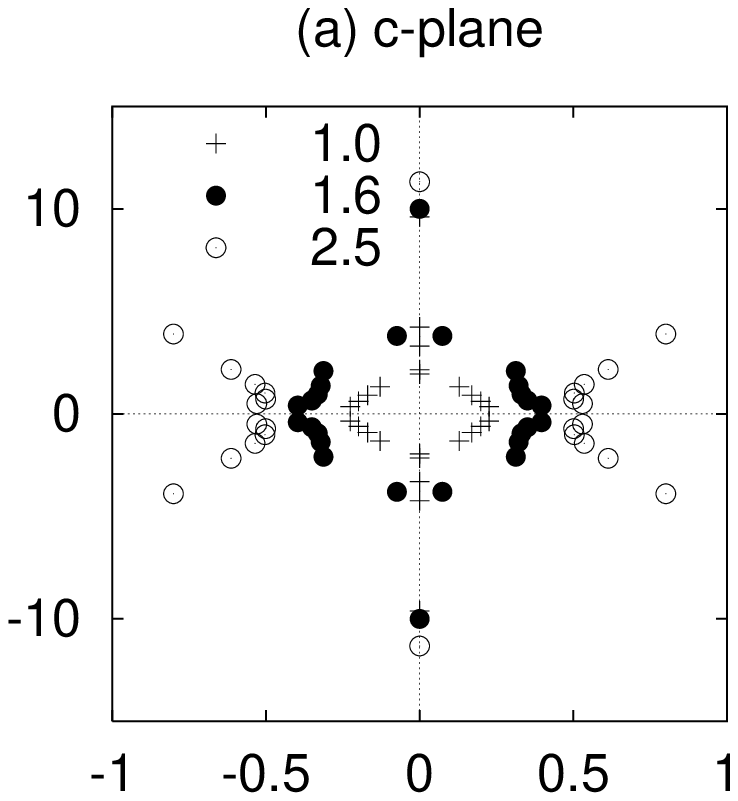}}
\end{minipage}
\hspace{-2.85cm}
\begin{minipage}[t]{6.5cm}
\centerline{\epsfxsize=6.5cm \epsfysize=4.33cm \epsfbox{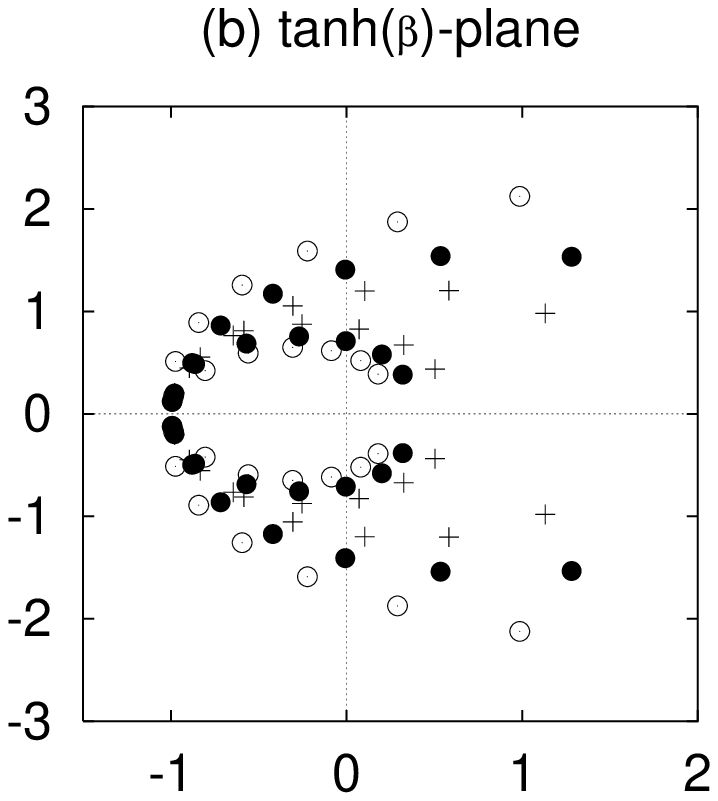}}
\end{minipage}}
 
\vspace{-2mm} 
\noindent{\bf Figure 2(a).}   
Fisher zeros in the complex $c$ plane for fixed $N=14$ and varying $y$. 
Only for $y=1$ ($H=0$) the zeros pinch the real axis. 
{\bf (b)} The trajectories in 
the ${\tilde c}$-plane. Note the FM and AFM phase transitions at 
$\tilde c=3/5,\ 5/3$ for $y=1$. The approach to these points is observed 
as $N$ increases (not shown).

\vspace{1mm}
{\it A priori}, it is not evident that the partition function zeros 
will exhibit scaling in the presence of quantum gravity.
However, the evidence in the range of central charge $0<c<1$
\cite{corr} for a diverging correlation length 
at the critical point makes such a scenario more plausible. 

From scaling arguments \cite{IPZ} 
one deduces that the $j$th  Lee--Yang zero should scale as 
$H_j \sim N^{-\beta\delta/(\nu d_H)}$.
Fig.~3 shows the first
three Lee--Yang zeros vs. $N$.
Fig.~3 gives ${\beta\delta/(\nu d_H)}\approx 0.871 \pm 0.002,\ 0.935 \pm 
0.002$ and
$0.951 \pm 0.002$ for $j=1,2,3$ respectively. The value expected from 
the scaling assumption is $5/6\approx 0.8333...$ for all $j$. 

It is expected from renormalization group arguments \cite{IPZ} 
that the trajectories of the motion of  
Fisher zeros in the $K$ plane ($K\equiv u-u_c$ 
and $u\equiv \e^{-4\beta}$) 
with varying real $H$, will form an angle $\psi = 
{{\pi}/{(2\beta\delta)}}$ 
with the real $K$ axis as $N\to \infty$.
For our case $\psi = 36^{\circ}$. The approach to this value 
(dashed line) can be observed in
Fig.~4. 

Next we tested the scaling relation \cite{IPZ}
$H_j^2(N/j)^{2\delta/(\delta+1)} =  F\left( K (N/j)^{1/(\nu d_H)}\right)$
where $F$ is a universal scaling function. 
Taking $K$ real, we plotted $H_j(N/j)^{5/6}$
vs. $(\beta-\beta_c)(N/j)^{1/3}$ (not shown)
for all zeros of order $j>1$ and all lattice sizes. 
The points fell on one universal 
curve representing the function $F$ for large $j$ in the hot phase.   
By varying the value of the exponent $5/6$ 
we observed a broadening of the curve and 
obtained a (somewhat subjective) determination 
${\delta/{(\delta+1)}} = {{\beta\delta}/{(\nu d_H)}} \approx 0.85 \pm 0.05$,
in excellent agreement with the KPZ value.

From the definition of $F$ we expect 
for the $j$th Fisher zero and large $N$: 
$\ln |K_j| \sim \ln |\beta_j - \beta_c| = 1/(\nu d_H) \ln (j/N) + C$
Fig.~5 shows $|K_1|$ and $|\beta_1 - \beta_c|$ vs. N on a log--log plot.
$K_j$ did not yield a straight line at these lattice sizes, but 
$|\beta_j-\beta_c|$ scales and gives $1/(\nu d_H)\approx 0.327 \pm 
0.001$,  $0.377 \pm 0.002$, for $j=1,2$ respectively. 
The value expected from scaling is $1/3$.

{\epsfxsize=6.0cm \epsfysize=4.0cm \epsfbox{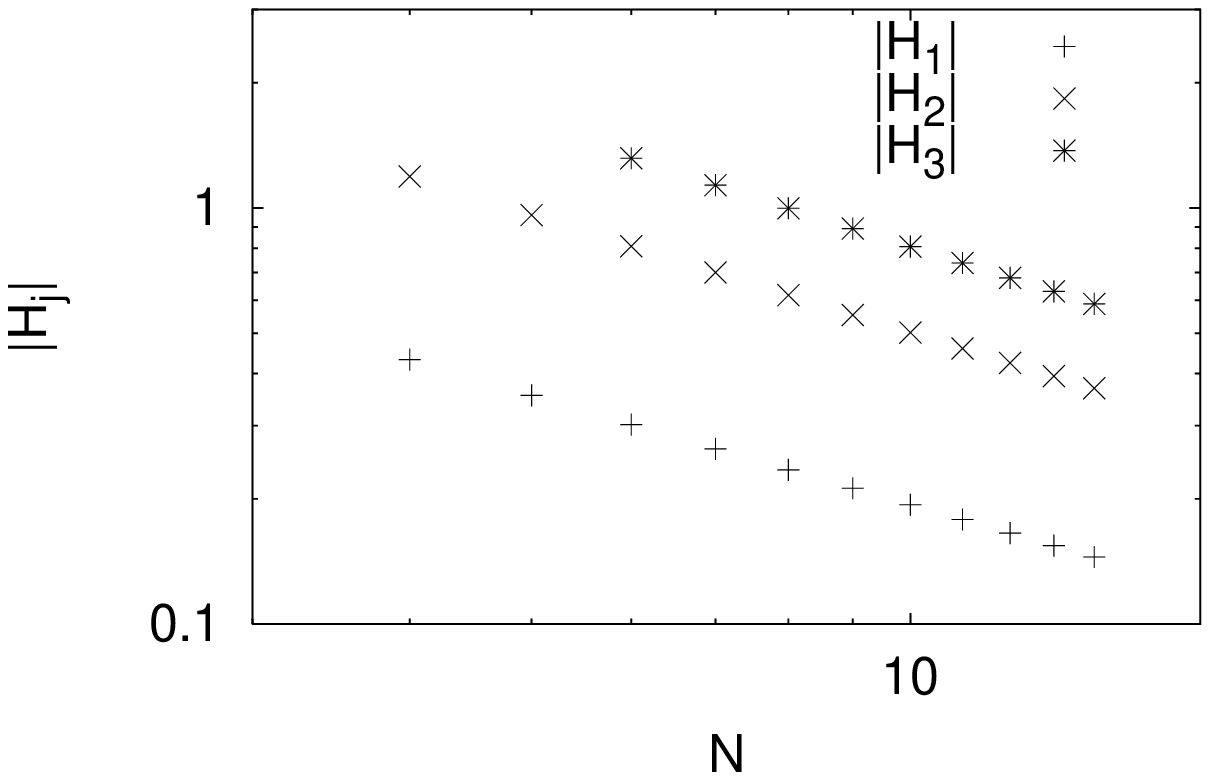}}

\vspace{-3.5mm}
\noindent {{\bf Figure 3.} Scaling of Lee--Yang zeros $H_j$.}

\vspace{1mm}
\centerline{\epsfxsize=5.0cm \epsfysize=3.33cm \epsfbox{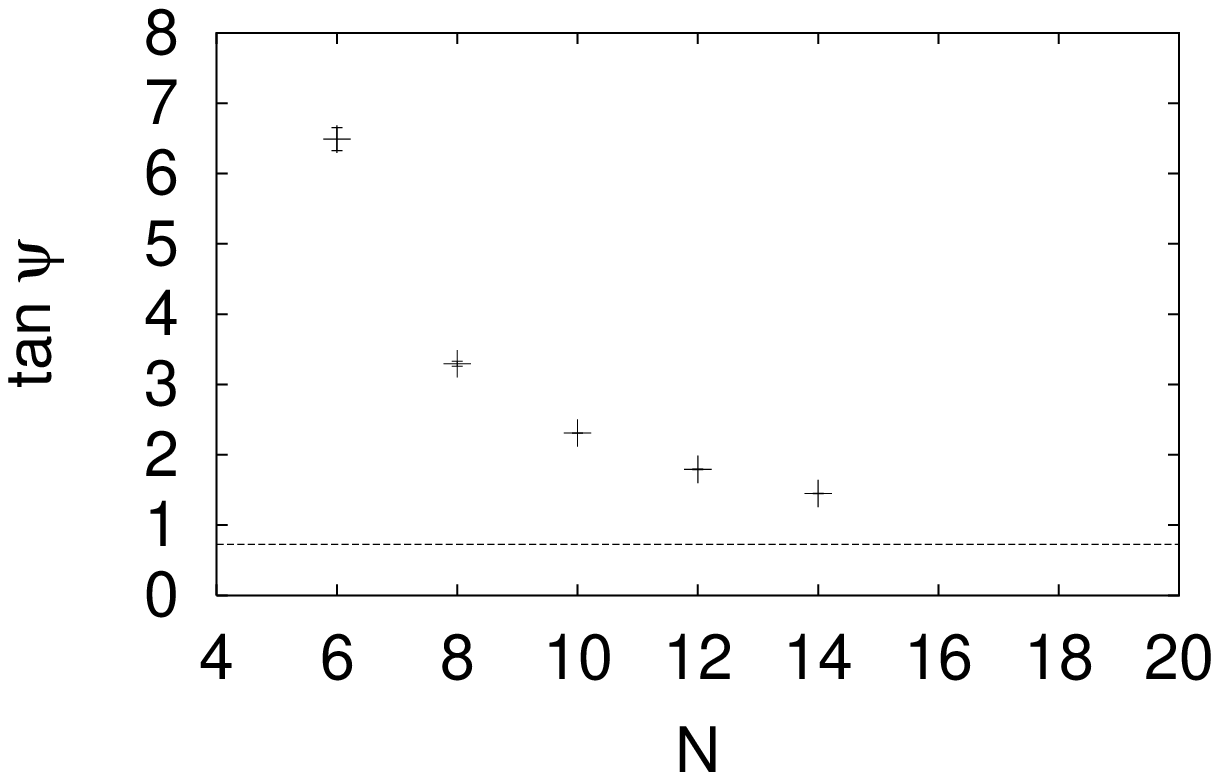}}

\vspace{-3mm}
\noindent {\bf Figure 4.} ${\rm tan}\psi$ versus $N$.
The dashed line is ${\rm tan}36^{\circ}$.

\vspace{1mm}
{\epsfxsize=6.0cm \epsfysize=4.0cm \epsfbox{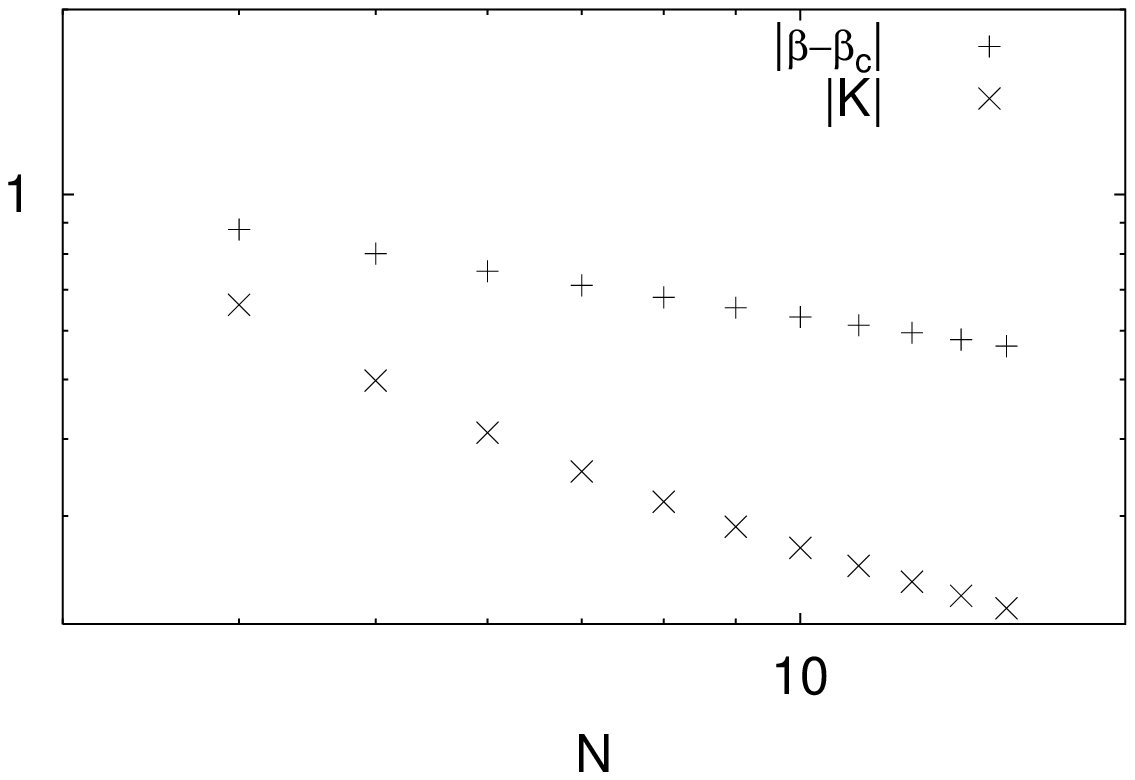}}

\vspace{-3.5mm}
\noindent {\bf Figure 5.} Scaling of the first Fisher zero.

\section{MULTIHISTOGRAMMING}

For systems with more than a few spins, the study of complex zeros
must rely on numerical methods.
We used multihistogramming \cite{MC} to 
determine the partition function  
accurately throughout a continuous range of couplings. 
The lattice sizes that we simulated ranged from 32 to 256 vertices
(60 -- 508 triangles). We refer to \cite{paper} for details.
Lee--Yang and Fisher zeros were determined from the minima of
${Z(\beta,H)}/{Z(\beta)}$
and ${Z(\beta,0)}/{Z({\rm Re}\beta)}$, respectively,
for real $\beta$ and imaginary $H$ and  
for complex $\beta$ and $H=0$.
Tables with the observed zeros can be found in 
\cite{paper}. We observed Lee--Yang zeros as follows: for $N_v=64$, 
only the first zero could be seen, 
for $N_v=96$, the first two zeros, for $N_v=128$,
three zeros, for $N_v=256$, five zeros. For the Fisher zeros, only the first
zero was visible for all lattice sizes $N_v=32,\ 64,\ 96,\ 128,\ 256$.

{\epsfxsize=6.0cm \epsfysize=4.0cm \epsfbox{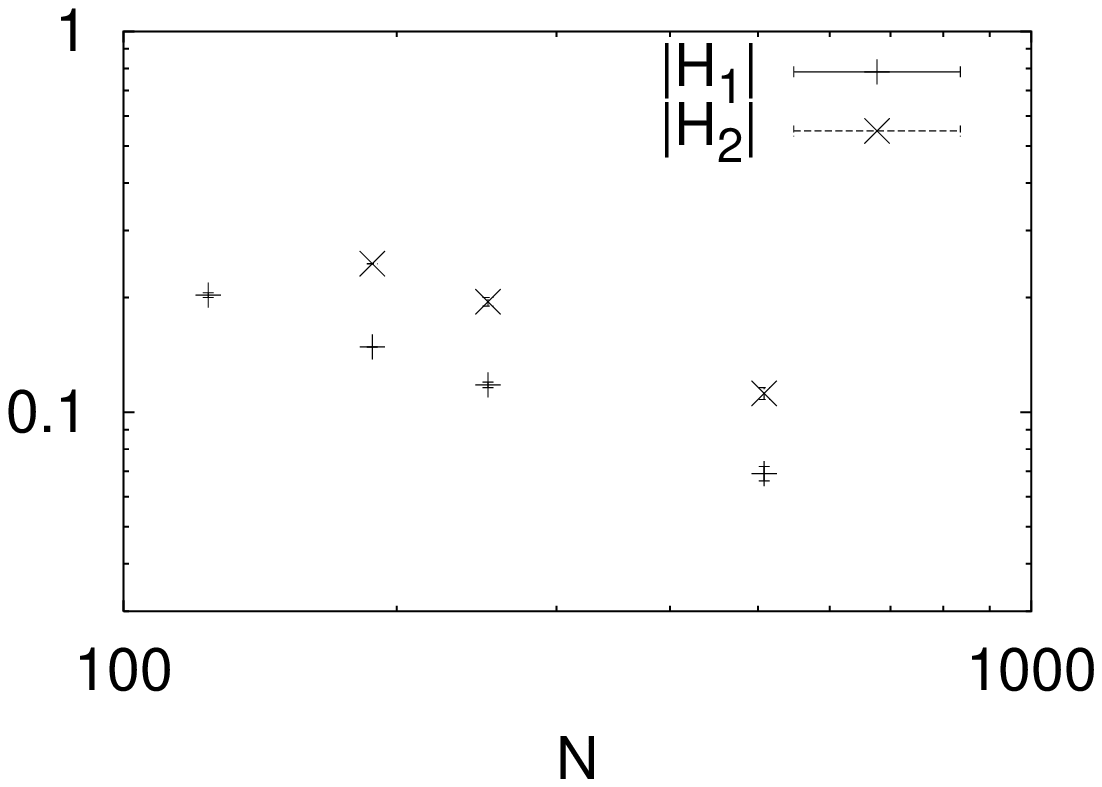}}

\vspace{-3mm}
\noindent {\bf Figure 6.} The 
first and second Lee--Yang zero observed
using multihistogramming.

{\epsfxsize=6.0cm \epsfysize=4.0cm \epsfbox{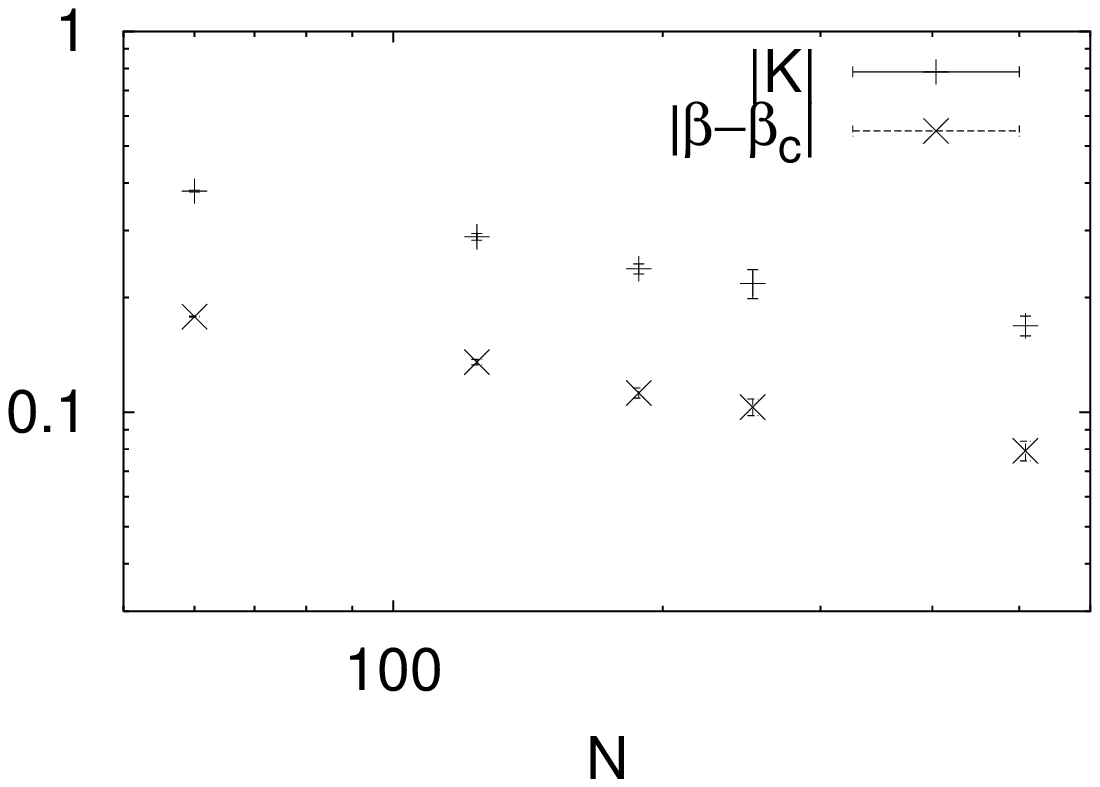}}

\vspace{-2mm}
\noindent {\bf Figure 7.} 
The first Fisher zero observed using multihistogramming.

\vspace{1mm}
Fits to $H_j \sim N^{-\beta\delta/(\nu d_H)}$ give for the
first and second Lee--Yang zero (Fig.~6)
${{\beta\delta}/{(\nu d_H)}} = 0.773 \pm 0.013,\ 0.788 \pm 0.033$ ($j=1,2$).
Using $N_v$ instead of $N$ for volume, the corresponding
values are 
$0.787 \pm 0.013,\ 0.800 \pm 0.034$ ($j=1,2$),
indicating finite size effects.  The
results are in quite good agreement with the expected value 5/6.
From $\nu d_H = 3$ \cite{BK} we obtain
$\beta\delta = 2.36 \pm 0.04,\ 2.40 \pm 0.10$ ($j=1,2$ respectively);
the exact value is 2.5.

The scaling of the observed Fisher zeros is shown in
Fig.~7. The extracted values of $1/(\nu d_H)$ are: for $K$ 0.392 $\pm$ 0.016;
for $|\beta-\beta_c|$ 0.386 $\pm$ 0.014 (expected value: 1/3 for both as 
$N\to \infty$). They increase slightly if 
$N_v$ is used instead of $N$.
Including only the three largest lattices we obtain (using $N$ for
volume): $1/{(\nu d_H)} = 0.350 \pm 0.067$.

\section{CONCLUSION}

We conclude that the critical behaviour of the system is
described by the scaling of the distribution of the complex zeros of the
partition function. Our result for Lee--Yang zeros 
presents us with the challenge of proving a corresponding Lee--Yang
theorem for the case when a fluctuating metric contributes an
additional quantum degree of freedom.

\end{document}